\documentclass[journal]{IEEEtran}

\usepackage{amsmath,graphicx}
\usepackage{comment}
\usepackage{cite}
\usepackage[utf8]{inputenc}
\usepackage{xcolor}
\usepackage{url}
\usepackage{amssymb}
\usepackage{subfigure}

\usepackage[linesnumbered,ruled,vlined,commentsnumbered]{algorithm2e}

\SetCommentSty{mycommfont}

\renewcommand{\v}[1]{\ensuremath{\mathbf{#1}}}
\DeclareMathOperator*{\argmax}{arg\,max}

\newcommand{\pdffigure}[3][width=0.7\linewidth]{
    \begin{figure}[tb]
    \begin{center}
    \IfFileExists{./#2.pdf}{
        \includegraphics[#1]{#2.pdf}
    }{
        \includegraphics[draft]{#2.pdf}
    }
    \end{center}
    \caption{#3}
    \label{fig:#2}
    \end{figure}
}

\begin{document}
%

\title{End-to-End Neural Diarization:\\ Reformulating Speaker Diarization as\\ Simple Multi-label Classification}

%
%
%

\author{
Yusuke~Fujita,~\IEEEmembership{Member,~IEEE},
Shinji~Watanabe,~\IEEEmembership{Senior~Member,~IEEE},
Shota~Horiguchi,~\IEEEmembership{Member,~IEEE},
Yawen~Xue,
and~Kenji~Nagamatsu%
\thanks{Yusuke Fujita, Shota Horiguchi, Yawen Xue, and Kenji Nagamatsu are with Research \& Development Group, Hitachi, Ltd., Japan,  e-mail:\{yusuke.fujita.su, shota.horiguchi.wk, yawen.xue.wn, kenji.nagamatsu.dm\}@hitachi.com.}
\thanks{Shinji Watanabe is with Johns Hopkins University, USA, e-mail: \mbox{shinjiw@ieee.org}.}
}

%
%

\markboth{IEEE/ACM TRANSACTIONS ON AUDIO, SPEECH, AND LANGUAGE PROCESSING,~Vol.~XX, No.~X, Xxxxx~2019}%
{Shell \MakeLowercase{\textit{et al.}}: Bare Demo of IEEEtran.cls for IEEE Journals}
%



\maketitle

\begin{abstract}
The most common approach to speaker diarization is clustering of speaker embeddings.
However, the clustering-based approach has a number of problems; i.e., (i) it is not optimized to minimize diarization errors directly, (ii) it cannot handle speaker overlaps correctly, and (iii) it has trouble adapting their speaker embedding models to real audio recordings with speaker overlaps.
To solve these problems, we propose the End-to-End Neural Diarization (EEND), in which a neural network directly outputs speaker diarization results given a multi-speaker recording.
To realize such an end-to-end model, we formulate the speaker diarization problem as a multi-label classification problem and introduce a permutation-free objective function to directly minimize diarization errors.
Besides its end-to-end simplicity, the EEND method can explicitly handle speaker overlaps during training and inference.
Just by feeding multi-speaker recordings with corresponding speaker segment labels, our model can be easily adapted to real conversations.
We evaluated our method on simulated speech mixtures and real conversation datasets.
The results showed that the EEND method outperformed the state-of-the-art x-vector clustering-based method, while it correctly handled speaker overlaps.
We explored the neural network architecture for the EEND method, and found that the self-attention-based neural network was the key to achieving excellent performance.
In contrast to conditioning the network only on its previous and next hidden states, as is done using bidirectional long short-term memory (BLSTM), self-attention is directly conditioned on all the frames.
By visualizing the attention weights, we show that self-attention captures global speaker characteristics in addition to local speech activity dynamics, making it especially suitable for dealing with the speaker diarization problem.

\end{abstract}


%
\IEEEpeerreviewmaketitle

\section{Introduction}
%
%
%
%
\label{sec:intro}
\IEEEPARstart{S}{peaker} diarization is the process of partitioning an audio recording into homogeneous segments according to the speaker's identity. Speaker diarization has a wide range of applications, such as generating written records of meetings and a turn-taking analysis of telephone conversations \cite{Tranter2006, Anguera2012}.
It also improves automatic speech recognition performance in multi-speaker conversation scenarios in meetings (ICSI \cite{Janin03, etin2006OverlapIM}, AMI \cite{Renals2008, Kanda2019ICASSP}) and home environments (CHiME-5 \cite{Barker2018, Du2018, Boeddecker2018, Kanda2018, Kanda2019ICASSP}).

The most common approach to speaker diarization is based on clustering of speaker embeddings \cite{Meignier2010LIUMSA, Shum2013, Sell2014, Senoussaoui2014, Dimitriadis2017, Romero2017, Maciejewski2018CharacterizingPO, Wang2018LSTM}.
For instance, i-vectors \cite{Dehak2011, Shum2013, Sell2014, Maciejewski2018CharacterizingPO}, d-vectors\cite{Wan2018, Wang2018LSTM}, and x-vectors \cite{Snyder2018, Romero2017} are commonly used in speaker diarization tasks.
These embeddings of short segments are partitioned into speaker clusters by using clustering algorithms, such as Gaussian mixture models \cite{Meignier2010LIUMSA, Shum2013}, agglomerative hierarchical clustering \cite{Meignier2010LIUMSA, Sell2014, Romero2017, Maciejewski2018CharacterizingPO},
mean shift clustering \cite{Senoussaoui2014}, k-means clustering \cite{Dimitriadis2017, Wang2018LSTM}, Links \cite{Mansfield2018,Wang2018LSTM}, and spectral clustering \cite{Wang2018LSTM}.
These clustering-based diarization methods have shown themselves to be effective on various datasets (see the DIHARD Challenge 2018 activities, e.g.,  \cite{Sell2018dihard, Diez2018, Sun2018}).

However, such clustering-based methods have a number of problems.
First, they cannot be optimized to minimize diarization errors directly because the clustering is a type of unsupervised learning process.
Second, they have trouble handling speaker overlaps, since the clustering algorithms implicitly assume one speaker per segment.
Furthermore, they have trouble adapting their speaker embedding models to real audio recordings with speaker overlaps, because the speaker embedding model has to be optimized with single-speaker non-overlapping segments.
These problems hinder speaker diarization when it is applied to real audio recordings that usually contain speaker overlaps.

To solve these problems, we propose End-to-End Neural Diarization (EEND).
Different from most of the other methods, EEND does not rely on clustering.
Instead, a neural network directly outputs the joint speech activities of all speakers for each time frame, given an input of a multi-speaker audio recording.
Our method can naturally handle speaker overlaps during the training and inference period by exploiting a multi-label classification framework.

EEND is based on our previous studies \cite{Fujita2019E2EDiarization, Fujita2019ASRU}.
In \cite{Fujita2019E2EDiarization}, we proposed an optimal training scheme for a diarization model with a permutation-free objective function that provides minimal diarization errors.
In \cite{Fujita2019ASRU}, we extended that method by exploiting a self-attention-based neural network.
Instead of a bidirectional long short-term memory (BLSTM) \cite{Graves2005}, we used a self-attention mechanism \cite{Lin2017, Vaswani2017}, which resulted in a significant performance improvement over the BLSTM-based model.

In this paper, we reformulate speaker diarization as a simple multi-label classification, which is independent of the choice of neural network architecture: BLSTM or self-attention. Then we investigate the proposed method from various perspectives, by comparison the effect of different network architectures, visualizing latent representations, and evaluating it on multiple real datasets.
The comparison of the network architectures revealed that a multi-head self-attention-based neural network is the key to achieving excellent performance.
Experiments with different numbers of heads showed that the excellent performance could be obtained by making the number of heads sufficiently larger than the number of speakers.
Experiments with different numbers of self-attention-based encoder blocks revealed that the EEND model performed better when it had more encoder blocks.
By visualizing the latent representation, we showed that self-attention could capture global speaker characteristics in addition to local speech activity dynamics, making it especially suitable for dealing with the speaker diarization problem.
The evaluation on the real datasets showed that the EEND method outperformed the state-of-the-art x-vector clustering-based method, while it correctly handled the speaker overlaps.

\section{Related work}
\label{sec:relatedwork}

\subsection{Clustering-based methods}

\begin{figure}[tb]
  \begin{center}
  \subfigure[Clustering-based method]{
    \includegraphics[width=0.97\linewidth]{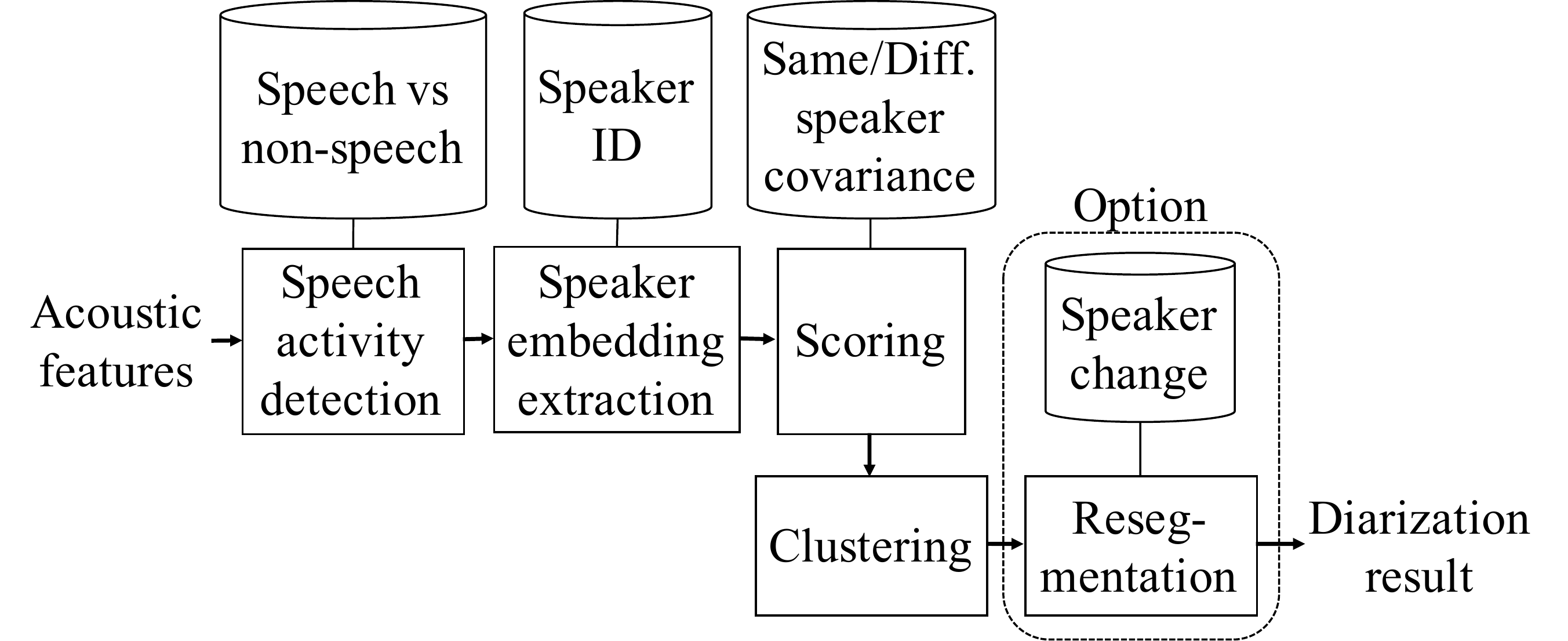}
    \label{fig:x-vector}
  }
  \subfigure[EEND method]{
    \includegraphics[width=0.7\linewidth]{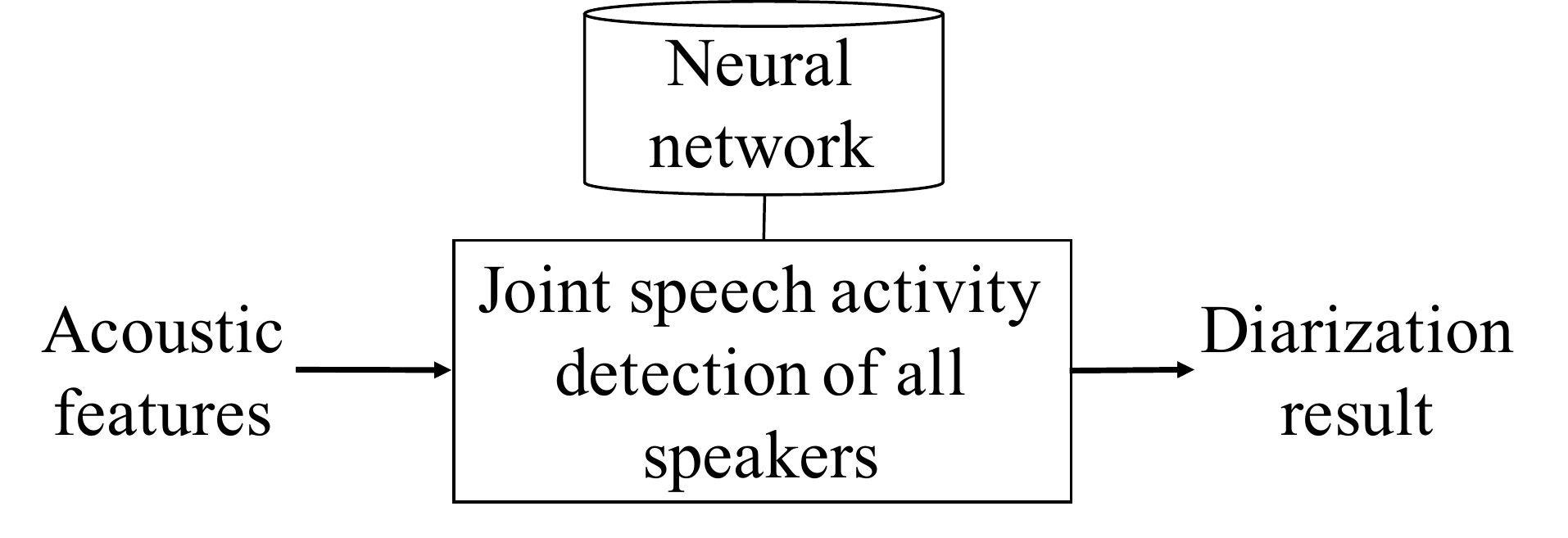}
    \label{fig:eend}
  }
  \end{center}
  \caption{System diagrams for speaker diarization. While clustering-based method requires three different models, EEND method requires one model.}
\label{fig:systems}
\end{figure}

Clustering-based methods are commonly used for speaker diarization. We used i-vector/x-vector clustering-based systems \cite{Sell2018dihard, Diez2018, Snyder2019} as the baselines in our experiments.
A diagram of a typical clustering-based system is depicted in Fig. \ref{fig:x-vector}.

To build the system, one has to prepare three independent models: (i) a speech activity detection (SAD) model for discriminating speech and non-speech, (ii) a speaker embedding extraction model for speaker identification, and (iii) a scoring model including the same/different speaker covariance matrices.
None of these models can be trained to minimize the diarization errors directly.
Optionally, a resegmentation process requires another model to refine speaker change points to produce the final diarization results.

Joint modeling methods have been studied in an effort to alleviate the complex preparation process and take into account the dependencies between these models.
They include, for example, joint modeling of speaker embedding extraction and scoring \cite{Romero2017, Narayanaswamy2019} and joint modeling of SAD and speaker embedding \cite{MiasatoFilho2018}.
However, the clustering process has remained unchanged because it is an unsupervised process.

In contrast to these methods, the EEND method uses only one neural network model, as depicted in Fig. \ref{fig:eend}. 
This method does not rely on clustering, and the model can be directly optimized with the reference diarization results of the training data.

This neural-network-based end-to-end approach, in which only one neural network model directly computes the final outputs, has been successfully applied in a variety of tasks, including neural machine translation \cite{Bahdanau2015, Sutskever2014}, automatic speech recognition \cite{Chorowski2015, Chan2016, Watanabe2017}, and text-to-speech \cite{Wang2017, Sotelo2017}. The proposed method is also categorized as such an approach.

\subsection{Clustering-free methods}

The clustering process impares the model optimization aimed at minimizing diarization errors. To alleviate this problem, Zhang et al. proposed a clustering-free diarization method \cite{Zhang2018}. 
This method is the first successful approach that does not cluster speaker embeddings and that is optimized with a diarization error minimization objective. 
The method formulates the speaker diarization problem on the basis of a factored probabilistic model, which consists of modules for determining speaker changes, speaker assignments, and feature generation.
These models are jointly trained using input features and corresponding speaker labels. However, the SAD model and their speaker embedding (d-vector) extraction model have to be trained separately in their method. Moreover, their speaker-change model assumes one speaker for each segment, which hinders its application to speaker-overlapping speech.

In contrast to their method, the EEND method uses an end-to-end neural network that accepts audio features as input and outputs the joint speech activities of multiple speakers. The network is optimized using the entire recording, including non-speech and speaker overlaps, with a diarization-error-oriented objective.

\subsection{Self-attention mechanism}

The self-attention mechanism was originally proposed for extracting sentence embeddings for text processing \cite{Lin2017}. Recently, the self-attention mechanism has shown superior performance in a variety of tasks, including machine translation \cite{Vaswani2017}, video classification \cite{Wang2018NonlocalNN}, and image segmentation \cite{Ye_2019_CVPR}. For audio processing, a self-attention mechanism has been incorporated in acoustic modeling for ASR \cite{Sperber2018, Dong2018}, sound event detection \cite{Wang2018}, and speaker recognition \cite{Zhu2018}. For speaker diarization, the self-attention mechanism has been applied to the speaker embedding extraction model \cite{Sun2018} and the scoring model \cite{Narayanaswamy2019} of clustering-based methods. This study describes a self-attention mechanism for clustering-free speaker diarization.

\section{End-to-End Neural Diarization (EEND)}
\label{sec:method}

In this section, we describe a novel approach to speaker diarization problem exploiting a multi-label classification framework with a permutation-free training scheme. We refer to the proposed method as EEND.

\subsection{Speaker diarization as multi-label classification}
\label{sec:e2e}
The speaker diarization task can be formulated as a  probabilistic multi-label classification problem, as follows.

Given an observation sequence of length $T$, $X = (\v{x}_t \in \mathbb{R}^{F} \mid t=1,\cdots,T)$, from an audio signal, the speaker diaization problem is one of estimating the corresponding speaker label sequence $Y = (\v{y}_t \mid t=1,\cdots,T)$.
Here, $\v{x}_t$ is an $F$-dimensional observation feature vector at time index $t$.
Speaker label $\v{y}_t = [y_{t,c} \in \{0,1\} \mid c=1, \cdots, C]$ denotes a joint activity for multiple ($C$) speakers at time index $t$.
For example, $y_{t,c} = y_{t,c'} = 1\; (c \ne c')$ represent an overlap situation in which speakers $c$ and $c'$ are both present at time index $t$.
Thus, determining $Y$ is a sufficient condition to determine the speaker diarization information.

The most probable speaker label sequence $\hat{Y}$ is selected from among all possible speaker label sequences $\mathcal{Y}$, as follows:
\begin{equation}
\hat{Y} = \argmax_{Y \in \mathcal{Y}} P(Y|X).
\end{equation}
$P(Y|X)$ can be factorized using the conditional independence assumption as follows:
\begin{align}
    P(Y|X) &= \prod_{t} P(\v{y}_t | \v{y}_1, \cdots \v{y}_{t-1}, X), \\
    &\approx \prod_t P(\v{y}_t| X) \approx \prod_t \prod_c P(y_{t,c}| X).
\end{align}
Here, we assume that the frame-wise posterior is conditioned
on all inputs, and each speaker is present independently.
The frame-wise posteriors can be estimated using a neural-network-based model, as follows:
\begin{equation}
    \mathbf{z}_t = [P(y_{t,1}|X), \cdots, P(y_{t,C}|X)] = \mathrm{NN}_t(X)  \in (0,1)^{C}, \label{eq:nnout}
\end{equation}
where $\mathrm{NN}_t(\cdot)$ is a neural network which accepts a sequence of input features and outputs $\mathbf{z}_t$, a $C$-dimensional vector of the frame-wise posteriors at time index $t$.

\subsection{Permutation-free training}

The difficulty of training the model described above is that the model must deal with speaker permutations: changing the order of speakers within a correct label sequence is also regarded as correct. An example of permutations in a two-speaker case is shown in Fig. \ref{fig:permutation-free}.
In this paper, we call this problem ``label ambiguity.'' This label ambiguity obstructs the training of the neural network when we use a standard binary cross-entropy loss function.

\pdffigure[width=\linewidth]{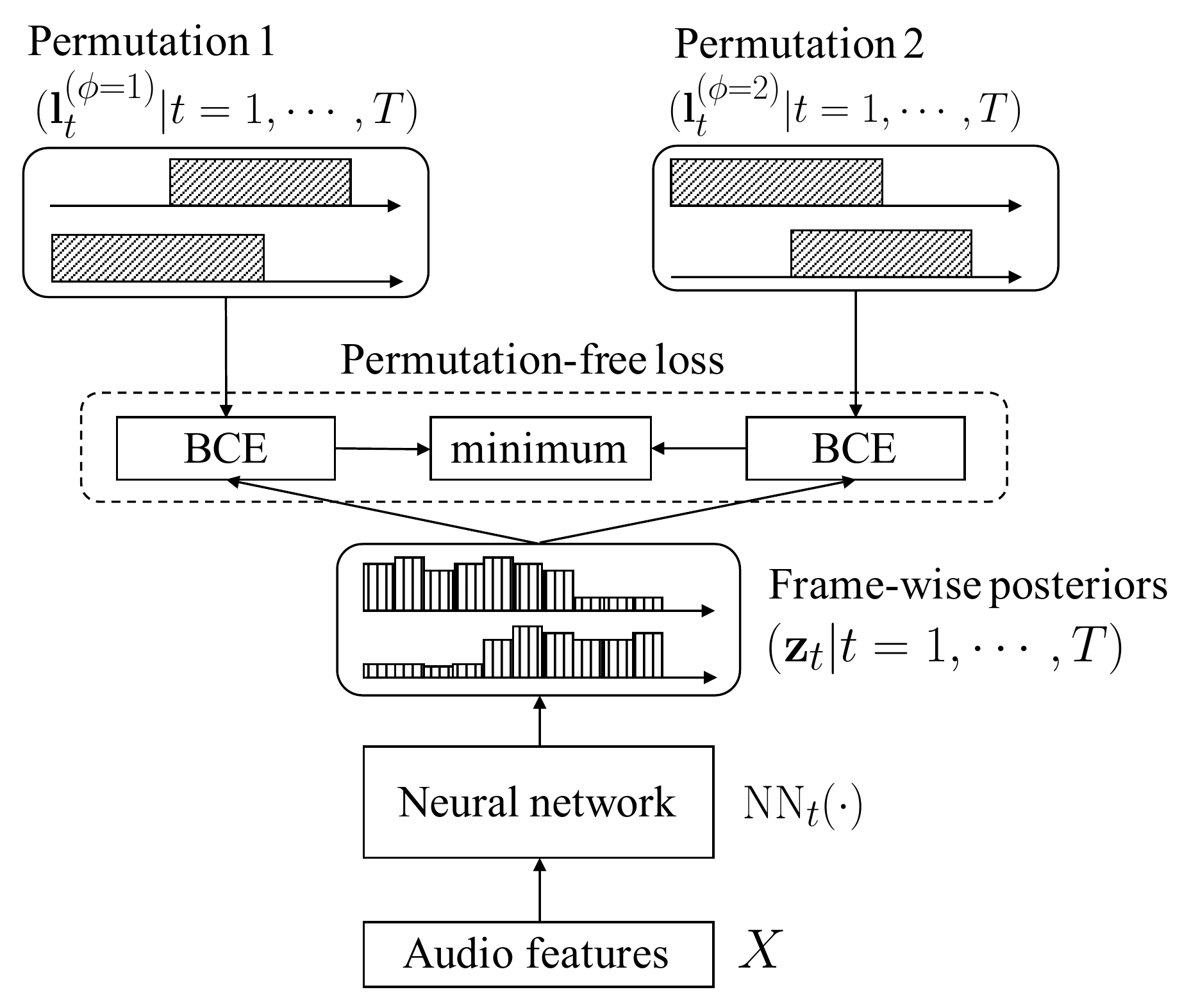}{Two-speaker EEND model trained with permutation-free loss. The binary cross entropy (BCE) loss of frame-wise posteriors $(\mathbf{z}_t \mid t=0,\cdots,T)$ are computed with two permutations of reference labels.}

To cope with the label ambiguity problem, we employ the permutation-free training scheme, which considers all the permutations of the reference speaker labels.
The permutation-free training scheme has been used in research on source separation \cite{Hershey2016, Yu2017, Kolbak2017}. Here, we apply a permutation-free loss function to a temporal sequence of speaker labels.
The neural network is trained to minimize the permutation-free loss between the output $\v{z}_t$ predicted using Eq. \ref{eq:nnout} and the reference speaker label $\v{l}_t \in \{0,1\}^C $, as follows:
\begin{equation}\label{eq:pf}
    J^{\text{PF}} = \frac{1}{TC} \min_{\phi \in \mathrm{perm}(C)} \sum_t \mathrm{BCE}(\v{l}_t^\phi, \v{z}_t),
\end{equation}
where $\mathrm{perm}(C)$ is the set of all the possible permutations of ($1,\dots,C$),
and $\v{l}_t^\phi$ is the $\phi$-th permutation of the reference speaker label, and $\mathrm{BCE}(\cdot, \cdot)$ is the binary cross entropy function between the label and the output.

\section{Neural network architectures for EEND}

In this section, we explore two different architectures of neural networks for the EEND method.

\subsection{BLSTM-based neural network with Deep Clustering loss}
\label{sec:blstm}

\pdffigure[width=\linewidth]{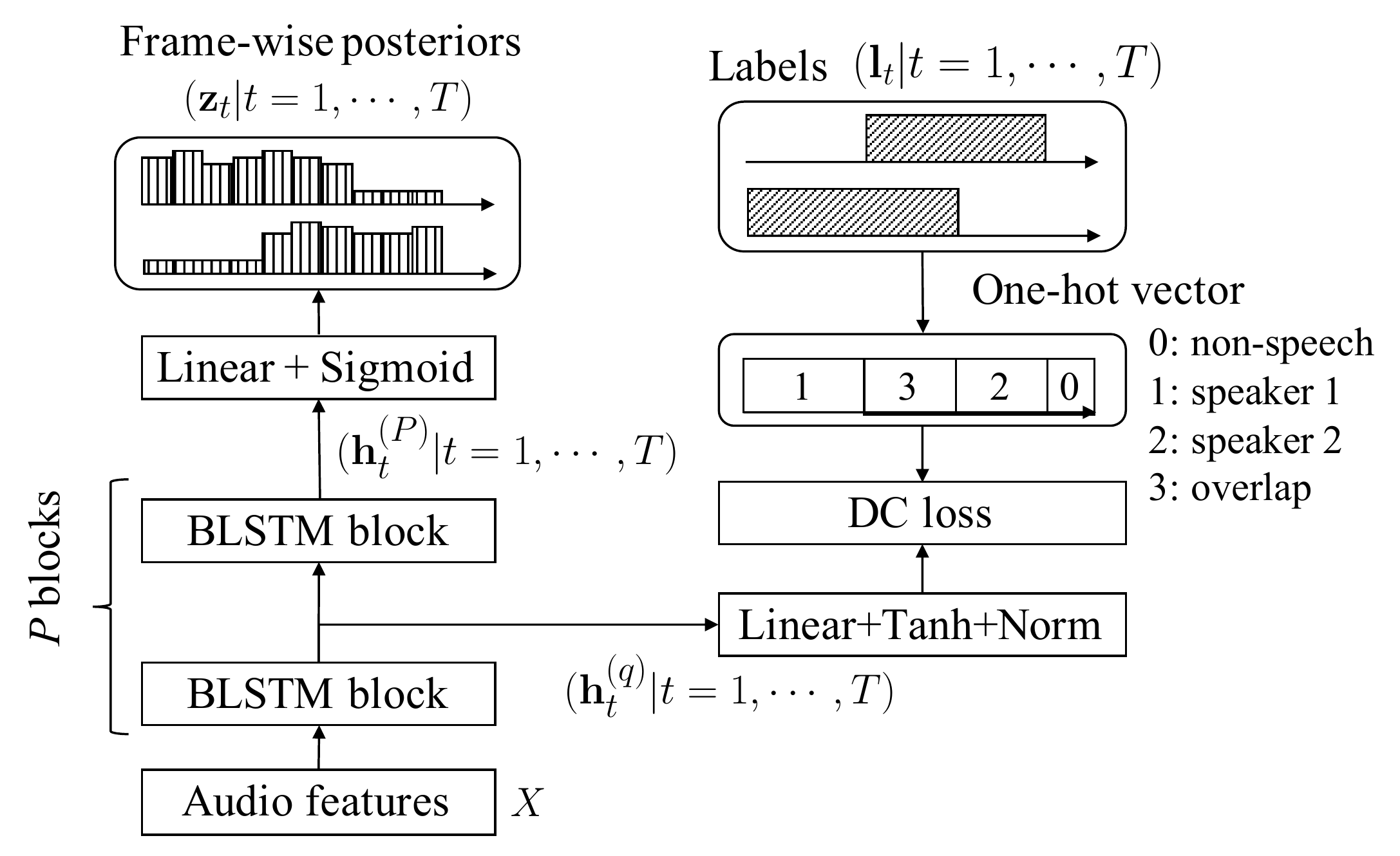}{BLSTM-based EEND model with Deep Clustering (DC) loss.}

According to Eq. \ref{eq:nnout}, the neural-network-based function $\mathrm{NN}_t(\cdot)$ accepts a temporal sequence of feature vectors and outputs a vector for each time frame. Thus, this function can be modeled with bi-directional long short-term memory (BLSTM) as depicted in Fig. \ref{fig:blstm}.
The input features are transformed as follows:
\begin{align}
    \v{h}^{(1)}_t &= \mathrm{BLSTM}_t^{(1)}(\v{x}_1, \cdots, \v{x}_T),\\
    \v{h}^{(p)}_t &= \mathrm{BLSTM}^{(p)}_t(\v{h}^{(p-1)}_1, \cdots, \v{h}^{(p-1)}_T) \  (2 \le  p \le P), \label{eq:blstmhidden} \\
    \v{z}_t  &= \sigma(\mathbf{W} \v{h}^{(P)}_t + \mathbf{b}),    \label{eq:blstmout}
\end{align}
where $\mathrm{BLSTM}^{(p)}_t(\cdot)$ is a $p$-th BLSTM layer which accepts an input sequence and outputs hidden activations $\v{h}_t^{(p)} \in\mathbb{R}^{2H}$ at time index $t$.\footnote{It is a concatenated vector of $H$-dimensional forward and backward LSTM outputs.}
$\mathbf{W} \in \mathbb{R}^{C \times 2H}$ and $\mathbf{b} \in \mathbb{R}^{C}$ are a linear projection matrix and bias, respectively.
We use $P$-layer stacked BLSTMs.

Assuming that the neural network extracts speaker embeddings in lower layers and then performs temporal segmentation using higher layers, the middle layer activations can be regarded as the speaker embeddings.
Therefore, we place a speaker embedding training criterion on the middle layer activations.

Here, the $q$-th layer activations $\v{h}^{(q)}_t$ obtained from Eq.~\ref{eq:blstmhidden} are transformed into normalized  $V$-dimensional embedding $\v{v}_t$ as follows:
\begin{align}
    \v{v}_t &= \mathrm{Normalize}(\mathrm{Tanh}(\mathbf{W}^{(\mathrm{DC})} \v{h}^{(q)}_t + \mathbf{b}^{(\mathrm{DC})})) \in \mathbb{R}^V \label{eq:dc3},
\end{align}
where $\mathbf{W}^{(\mathrm{DC})} \in \mathbb{R}^{V \times 2H}$ and $\mathbf{b}^{(\mathrm{DC})} \in \mathbb{R}^V$ are a linear projection matrix and a bias, respectively. $\mathrm{Tanh}(\cdot)$ is the element-wise hyperbolic tangent function and $\mathrm{Normalize}(\cdot)$ is the L2 normalization function.
We apply the Deep Clustering (DC) loss function \cite{Hershey2016} so that the embeddings are partitioned into speaker-dependent clusters as well as overlapping and non-speech clusters.
For example, in a two-speaker case, we generate four clusters (Non-speech, Speaker 1, Speaker 2, and Overlapping) as shown in Fig. \ref{fig:blstm}.

DC loss function is expressed as follows:
\begin{align}
    J^\text{DC} &= \|\mathbf{V}\mathbf{V}^\top - \mathbf{L}'\mathbf{L}'^\top\|_F^2,
\end{align}
where $\mathbf{V} = [\v{v}_1 \cdots \v{v}_T]^\top$, and $\mathbf{L}' \in \mathbb{R}^{T \times 2^C}$ is a matrix in which each row represents a one-hot vector converted from $\v{l}_t$, where those elements are in the power set of speakers.
$\|\cdot\|_F$ is the Frobenius norm.
The loss function encourages the two embeddings at different time indices to be close together if they are in the same cluster and far away if they are in different clusters.

Next, we use multi-objective training introducing a mixing parameter $\alpha$:
\begin{align}
    J^{\text{MULTI}} = (1 - \alpha) J^{\text{PF}} + \alpha J^{\text{DC}}.
\end{align}

\subsection{Self-attention-based neural network}
\label{sec:sa}
\pdffigure[width=\linewidth]{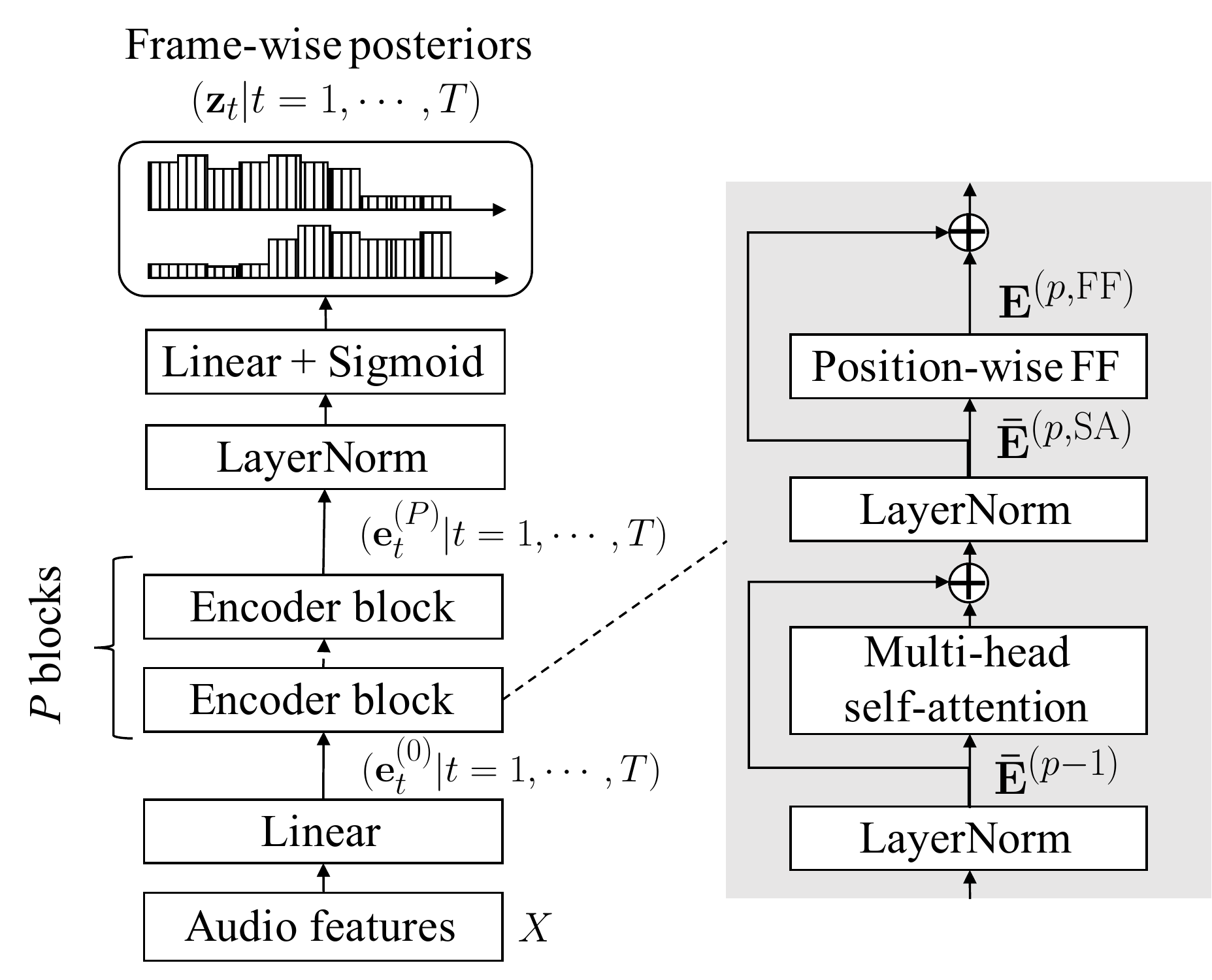}{Self-attention-based EEND model.}
By using BLSTM, each output frame is conditioned only on its previous hidden state, subsequent hidden state and current input feature. In contrast, by using a self-attention mechanism \cite{Lin2017}, each output frame is directly conditioned on all input frames by computing the pairwise similarity between all frame pairs.
Here, we use a self-attention-based neural network instead of BLSTM, as depicted in Fig. \ref{fig:self-att}.
The input features are transformed as follows:
\begin{align}
    \v{e}^{(0)}_t &= \mathbf{W}_0\v{x}_t + \mathbf{b}_0 \in \mathbb{R}^{D}, \\
    \v{e}^{(p)}_t &= \mathrm{Encoder}^{(p)}_t(\v{e}^{(p-1)}_1, \cdots, \v{e}^{(p-1)}_T) \  (1 \le  p \le P). \label{eq:hidden}
\end{align}
Here,  $\mathbf{W}_0 \in \mathbb{R}^{D \times F}$ and $\mathbf{b}_0 \in \mathbb{R}^D $ project an input feature into $D$-dimensional vector.
$\mathrm{Encoder}^{(p)}_t(\cdot)$ is the $p$-th encoder block which accepts an input sequence of $D$-dimensional vectors and outputs a $D$-dimensional vector $\v{e}_t^{(p)}$ at time index $t$. We use $P$ encoder blocks followed by the output layer for frame-wise posteriors.

The detailed architecture of the encoder block is depicted in Fig. \ref{fig:self-att}.
This configuration of the encoder block is almost the same as the one in the Speech-Transformer introduced in \cite{Dong2018}, but without positional encoding.
The encoder block has two sub-layers. The first is a multi-head self-attention layer, and the second is a position-wise feed-forward layer.
\subsubsection{Multi-head self-attention layer}
The multi-head self-attention layer transforms a sequence of input vectors as follows.
The sequence of vectors $(\mathbf{e}^{(p-1)}_t | t = 1, \cdots, T)$ is converted into a $\mathbb{R}^{T \times D}$ matrix; that is followed by layer normalization \cite{Ba2016}:
\begin{equation}
    \mathbf{\bar{E}}^\mathrm{(p-1)} = \mathrm{LayerNorm}([\mathbf{e}^{(p-1)}_1 \cdots \mathbf{e}^{(p-1)}_T]^\top) \in \mathbb{R}^{T \times D}. \label{eq:ln1}
\end{equation}
Then, query, key and value vectors are computed for each head $h\ (1\le h \le H)$ by using linear transformations:
\begin{align}
     \mathbf{Q}_h^{(p)} &= \mathbf{\bar{E}}^{(p-1)} \mathbf{W}^{(p,Q)}_{h} + \mathbf{1} \mathbf{b}^{(p,Q)\top}_{h} \in \mathbb{R}^{T \times d}, \\
     \mathbf{K}_h^{(p)} &= \mathbf{\bar{E}}^{(p-1)} \mathbf{W}^{(p,K)}_{h} + \mathbf{1} \mathbf{b}^{(p,K)\top}_{h}\in \mathbb{R}^{T \times d}, \\
     \mathbf{V}_h^{(p)} &= \mathbf{\bar{E}}^{(p-1)} \mathbf{W}^{(p,V)}_{h} + \mathbf{1}\mathbf{b}^{(p,V)\top}_{h} \in \mathbb{R}^{T \times d},
     \end{align}
where $d = D/H$ is the dimension of each head, $H$ is the number of heads, $\mathbf{W}^{(p,Q)}_h, \mathbf{W}^{(p,K)}_h,\mathbf{W}^{(p, V)}_h \in \mathbb{R}^{D \times d}$ are query, key, and value projection matrices, respectively. $\mathbf{b}^{(p,Q)}_{h}, \mathbf{b}^{(p,K)}_{h}, \mathbf{b}^{(p,V)}_{h} \in \mathbb{R}^d$ are bias vectors and $\mathbf{1}$ is a $T$-dimensional all-one vector.
A pairwise similarity matrix $\mathbf{A}^{(p)}_h$ is computed using the dot products of the query vectors and key vectors:
\begin{equation}
        \mathbf{A}^{(p)}_h =  \mathbf{Q}_h^{(p)} {\mathbf{K}_h^{(p)}}^\top \in \mathbb{R}^{T \times T}. \label{eq:a}
\end{equation}
The pairwise similarity matrix $\mathbf{A}^{(p)}_h$ is scaled by $1/\sqrt{d}$, and a softmax function is applied to form the attention weight matrix $\mathbf{\hat{A}}^{(p)}_h$:
\begin{equation}
        \mathbf{\hat{A}}^{(p)}_h = \mathrm{Softmax}\left(\frac{\mathbf{A}^{(p)}_h}{ \sqrt{d}}\right) \in \mathbb{R}^{T \times T}. \label{eq:softmax}
\end{equation}
Then, using the attention weight matrix, the context vectors $\mathbf{C}^{(p)}_h$ are computed as a weighted sum of the value vectors :
\begin{equation}
        \mathbf{C}^{(p)}_h = \mathbf{\hat{A}}^{(p)}_h \mathbf{V}^{(p)}_h \in \mathbb{R}^{T \times d}. \label{eq:value}
\end{equation}
Finally, the context vectors for all heads are concatenated and projected using an output projection matrix $\mathbf{W}^{(p,O)} \in \mathbb{R}^{D \times D}$ and a bias $\mathbf{b}^{(p,O)}$:
\begin{equation}
    \mathbf{E}^{(p,\mathrm{SA})} = [\mathbf{C}^{(p)}_1 \cdots \mathbf{C}^{(p)}_H] \mathbf{W}^{(p,O)} + \mathbf{1}\mathbf{b}^{(p,O)\top} \in \mathbb{R}^{T \times D}. \label{eq:concat}
\end{equation}
Following the self-attention layer, a residual connection and layer normalization are applied:
\begin{equation}
    \mathbf{\bar{E}}^{(p,\mathrm{SA})} = \mathrm{LayerNorm}(\mathbf{\bar{E}}^{(p-1)} + \mathbf{E}^{(p,\mathrm{SA})}) \in \mathbb{R}^{T \times D}. \label{eq:res}
\end{equation}

\subsubsection{Position-wise feed-forward layer}

The position-wise feed-forward layer transforms $\mathbf{\bar{E}}^{(p,\mathrm{SA})}$ as follows:
\begin{align}
    \mathbf{E}^{(p,\mathrm{ff})} &= \mathrm{ReLU}( \mathbf{\bar{E}}^{(p,\mathrm{SA})}\mathbf{W}^{(p)}_1 + \mathbf{1} \mathbf{b}^{(p)\top}_1 ) \in \mathbb{R}^{T \times d_{ff}}, \\
    \mathbf{E}^{(p,\mathrm{FF})} &= \mathbf{E}^{(p,\mathrm{ff})} \mathbf{W}^{(p)}_2 + \mathbf{1} \mathbf{b}^{(p)\top}_2 \in \mathbb{R}^{T \times D},
\end{align}
where $\mathbf{W}^{(p)}_1 \in \mathbb{R}^{D \times d_\mathrm{ff}}$ and $\mathbf{b}^{(p)}_1 \in \mathbb{R}^{d_\mathrm{ff}}$ are the first linear projection matrix and bias, respectively, and $\mathrm{ReLU}(\cdot)$ is the rectified linear unit activation function.
$d_\mathrm{ff}$ is the number of internal units in this layer. 
$\mathbf{W}^{(p)}_2 \in \mathbb{R}^{d_\mathrm{ff} \times D}$ and $\mathbf{b}^{(p)}_2 \in \mathbb{R}^{D}$ are the second linear projection matrix and bias, respectively.

Finally, the output of the encoder block $\mathbf{e}_t^{(p)}$ for each time frame is computed by applying a residual connection as follows:
\begin{equation}
    [\mathbf{e}_1^{(p)} \cdots \mathbf{e}_T^{(p)}] = (\mathbf{\bar{E}}^{(p,\mathrm{SA})} + \mathbf{E}^{(p,\mathrm{FF})})^\top
    \label{eq:resff}
\end{equation}

\subsubsection{Output layer for frame-wise posteriors}

The frame-wise posteriors $\mathbf{z}_t$ are calculated from $\mathbf{e}_t^{(P)}$ (in Eq. \ref{eq:hidden}) by using layer normalization and a fully-connected layer as follows:
\begin{align}
\bar{\mathbf{E}}^{(P)}&=\mathrm{LayerNorm}([\mathbf{e}_1^{(P)} \cdots \mathbf{e}_T^{(P)}]^\top) \in \mathbb{R}^{T\times D}, \\
[\mathbf{z}_1 \cdots \mathbf{z}_T] &= \sigma(\bar{\mathbf{E}}^{(P)} \mathbf{W}_3 + \mathbf{1} \mathbf{b}^\top_3 )^\top, \label{eq:out}
\end{align}
where $\mathbf{W}_3 \in \mathbb{R}^{D \times C}$ and $\mathbf{b}_3 \in \mathbb{R}^{C}$ are the linear projection matrix and bias, respectively, and $\sigma(\cdot)$ is the element-wise sigmoid function.

\section{Experimental setup}
\label{sec:setup}

\subsection{Data}
\label{sec:data}

\begin{table}[t]
\caption{Statistics of training and test sets.}
\label{tab:set}
\centering
\begin{tabular}{ll|rrr} \hline
    & & Num. of & Avg. dur. & \multicolumn{1}{c}{Overlap} \\
    & & mixtures & (sec) & ratio (\%) \\\hline
\multicolumn{2}{l|}{Traning sets} & & & \\
SimBeta2 & Simulated ($\beta=2$) & 100,000 & 87.6 & 34.4 \\
Real & SWBD+SRE & 26,172 & 304.7 & 3.7 \\
SimLarge &Simu. ($\beta=2,3,5,7$) & 400,000 & 126.4 & 23.4 \\
Comb & Real+SimLarge & 426,172 & 137.3 & 20.5 \\
\hline
\multicolumn{2}{l|}{Test sets} & & & \\
1& Simulated ($\beta=2$) & 500 & 87.3 & 34.4 \\
2& Simulated ($\beta=3$) & 500 & 103.8 & 27.2 \\
3& Simulated ($\beta=5$) & 500 & 137.1 & 19.5 \\
4& CALLHOME \cite{callhome} & 148 & 72.1 & 13.0 \\
5& CSJ \cite{Maekawa2003} & 54 & 766.3 & 20.1 \\ \hline
\end{tabular}
\end{table}

To verify the effectiveness of the EEND method for various overlap situations, we prepared four training sets and five test sets, including simulated and real datasets.
The statistics of the training and test sets are listed in Table \ref{tab:set}.
The overlap ratio is computed as the ratio of the audio time during which two or more speakers are active to the audio time during which one or more speakers are active.

Note that the training data for the EEND method are different from those for the i-vector/x-vector clustering-based method.
Whereas the clustering-based methods use single-speaker segments for training their speaker embedding extraction models, the EEND method uses audio mixtures of multiple speakers. Such mixtures can be simulated infinitely with a combination of single-speaker segments. Moreover, the EEND model can be trained with not only simulated mixtures but also real audio mixtures with speaker overlaps.

\subsubsection{Simulated datasets}


Each mixture was simulated by Algorithm \ref{alg:mixture_simulation}.
Unlike the mixture simulations of source separation studies \cite{Hershey2016}, we consider a diarization-style mixture: each speech mixture should have dozens of utterances per speaker with reasonable silence intervals between utterances.
The silence intervals are controlled by the average interval of $\beta$. Larger values of $\beta$ generate speech with less overlap.
\begin{algorithm}[t]
    \SetAlgoLined
    \DontPrintSemicolon
    \caption{Mixture simulation.}
    \label{alg:mixture_simulation}
    \SetAlgoVlined
    \SetKwInOut{Input}{Input}
    \SetKw{In}{in}
    \Input{
        {{$\mathcal{S,N,I,R}$} \tcp*{Sets of speakers, noises, RIRs and SNRs}}\\
        {{$\mathcal{U} = \{U_s\}_{s \in \mathcal{S}}$} \tcp*{Set of utterance lists}}
        {{$N_\text{spk}$} \tcp*{\#speakers per mixture}}
        {{$N_\text{umax},N_\text{umin}$} \tcp*{Max. and min. \#utterances per speaker}}
        {{$\beta$} \tcp*{Average interval}}
    }
    \SetKwInOut{Output}{Output}
    \Output{$\mathbf{y}$\tcp*{Mixture}}
    \BlankLine
    Sample a set of $N_\text{spk}$ speakers $\mathcal{S'}$ from $\mathcal{S}$\\
        $\mathcal{X}\leftarrow\emptyset$\tcp*{Set of $N_\text{spk}$ speakers' signals}
        \ForAll{$s\in\mathcal{S'}$}{
            $\mathbf{x}_s\leftarrow\emptyset$\tcp*{Concatenated signal}
            Sample $\mathbf{i}$ from $\mathcal{I}$\tcp*{RIR}
            Sample $N_u$ from $\left\{N_\text{umin},\dots,N_\text{umax}\right\}$
            
            \For{$u=1$ \rm{to} $N_u$}{
                Sample $\delta\sim\frac{1}{\beta}\exp\left(-\frac{\delta}{\beta}\right)$\tcp*{Interval}
                $\mathbf{x}_s\leftarrow\mathbf{x}_s\oplus\mathbf{0}^{\left(\delta\right)}\oplus U_s\left[u\right]\ast\mathbf{i}$
            }
            $\mathcal{X}.\mathsf{add}\left(\mathbf{x}_s\right)$\\
        }
        $L_\mathrm{max}=\max_{\mathbf{x}\in\mathcal{X}}\lvert\mathbf{x}\rvert$\\
        $\mathbf{y}\leftarrow\sum_{\mathbf{x}\in\mathcal{X}}\left(\mathbf{x}\oplus\mathbf{0}^{\left(L_\mathrm{max}-\lvert\mathbf{x}\rvert\right)}\right)$\\
        Sample $\mathbf{n}$ from $\mathcal{N}$\tcp*{Background noise}
        Sample $r$ from $\mathcal{R}$\tcp*{SNR}
        Determine a mixing scale $p$ from $r,\mathbf{y},$ and $\mathbf{n}$\\
        $\mathbf{n}'\leftarrow$ repeat $\mathbf{n}$ until the length of $\mathbf{y}$ is reached\\
        $\mathbf{y}\leftarrow\mathbf{y}+p\cdot\mathbf{n}'$\\
\end{algorithm}

The set of utterances used in the simulation was comprised of the Switchboard-2 (Phase I, II, III), Switchboard Cellular (Part 1, Part2), and NIST Speaker Recognition Evaluation datasets (2004, 2005, 2006, 2008). All recordings are telephone speech sampled at 8 kHz. There are 6,381 speakers in total. We split them into 5,743 speakers for the training set and 638 speakers for the test set.
Note that the set of utterances for the training set is identical to that of the Kaldi CALLHOME diarization v2 recipe \cite{Povey_ASRU2011}\footnote{\url{https://github.com/kaldi-asr/kaldi/tree/master/egs/callhome_diarization}}, thereby enabling a fair comparison with the x-vector clustering-based method.

Since there are no time annotations in these corpora, we extracted utterances using speech activity detection (SAD) on the basis of time-delay neural networks and statistics pooling\footnote{The SAD model: \url{http://kaldi-asr.org/models/m4}}.

The set of background noises was from the MUSAN corpus \cite{Snyder2015}. We used 37 recordings that are annotated as ``background'' noises.
The set of 10,000 room impulse responses (RIRs) was from the Simulated Room Impulse Response Database used in \cite{Ko2017}.
The SNR values were sampled from 10, 15, and 20 dB.
These sets of non-speech corpora were also used for training the x-vector and SAD models in the x-vector clustering-based method.

We generated two-speaker mixtures for each speaker with 10-20 utterances ($N_{\text{spk}} = 2, N_{\text{umin}}=10, N_{\text{umax}}=20$).
For the simulated training set, 100,000 mixtures were generated with $\beta=2$ (SimBeta2).
In addition, four sets of 100,000 mixtures with different values of $\beta$ (2, 3, 5, and 7) were combined to form 400,000 mixtures (SimLarge).
For the simulated test set, 500 mixtures were generated with $\beta=2$, 3, and 5.
The overlap ratios of the simulated mixtures ranged from 19.5 to 34.4\%.

\subsubsection{Real datasets}

We used real telephone speech recordings as the real training set (Real). A set of 26,172 two-speaker recordings were extracted from the recordings of the Switchboard-2 (Phase I, II, III), Switchboard Cellular (Part 1, Part 2), and NIST Speaker Recognition Evaluation datasets.
The overlap ratio of the training data was 3.7\%, far less than that of the simulated mixtures.

We evaluated the proposed method on real telephone conversations in the CALLHOME dataset \cite{callhome}.
We randomly split the two-speaker recordings from the CALLHOME dataset into two subsets: an adaptation set of 155 recordings and a test set of 148 recordings.
The average overlap ratio of the test set was 13.0\%.

In addition, we conducted an evaluation on the dialogue part of the Corpus of Spontaneous Japanese (CSJ) \cite{Maekawa2003}. The CSJ contains 54 two-speaker dialogue recordings\footnote{We excluded four out of 58 recordings that contain speakers in the official speech recognition evaluation sets.}. They were recorded using headset microphones in separate soundproof rooms. The average overlap ratio of the CSJ test set was 20.1\%, larger than the CALLHOME test set.

\subsubsection{Combined datasets}

For generalizing a model to various environments, we conducted experiments using both a simulated training set (SimLarge) and the real training set (Real). We refer to the dataset as the combined training set (Comb).

\subsection{Model configuration}
\subsubsection{Clustering-based systems}
We compared the proposed method with two conventional clustering-based systems \cite{Sell2018dihard}: the i-vector system and x-vector system were created using the Kaldi CALLHOME diarization v1 and v2 recipes.

These recipes use agglomerative hierarchical clustering (AHC) with the probabilistic linear discriminant analysis (PLDA) scoring scheme. The number of clusters was fixed to 2. Though the original recipes use oracle speech/non-speech marks, we used the SAD model with the configuration described in Sec. \ref{sec:data}.

\subsubsection{BLSTM-based EEND system}

We configured the BLSTM-based EEND system (BLSTM-EEND) described in Sec. \ref{sec:blstm}. 
The input features were 23-dimensional log-Mel-filterbanks with a 25-ms frame length and 10-ms frame shift. Each feature was concatenated with those from the previous seven frames and subsequent seven frames. To deal with a long audio sequence in our neural networks, we subsampled the concatenated features by a factor of ten. Consequently, a $(23\times 15)$-dimensional input feature was fed into the neural network every 100 ms.

We used a five-layer BLSTM with 256 hidden units in each layer. The second layer of the BLSTM outputs was used to form a 256-dimensional embedding; we then calculated the Deep Clustering loss in this embedding to discriminate different speakers. The mixing parameter $\alpha$ was set to 0.5. We used the Adam \cite{Kingma2014} optimizer with a learning rate of $10^{-3}$. The batch size was 10. The number of training epochs was 20.

Because the output of the neural network is the probability of speech activity for each speaker, a threshold is required to obtain a decision on speech activity for each frame. We set the threshold to 0.5. Furthermore, we applied 11-frame median filtering to prevent production of unreasonably short segments.

For domain adaptation, the neural network was retrained using the CALLHOME adaptation set.
We used the Adam optimizer with a learning rate of $10^{-6}$ and ran five epochs.
For the postprocessing, we adjusted the threshold to 0.6 so that the DER of the adaptation set had the minimum value.

\subsubsection{Self-attention-based EEND system}
\label{sec:eendsetup}
We configured a Self-attention-based EEND system (SA-EEND) as described in Sec. \ref{sec:sa}. 
Here, we used the same input features as were input to the BLSTM-EEND system. Note that the sequence length in the training stage was limited to 500 (50 seconds in audio time) because our system uses more memory than the BLSTM-based network does. Therefore, we split the input audio recordings into non-overlapping 50-second segments. In the inference stage, we used the entire sequence for each recording.

We used two encoder blocks with 256 attention units containing four heads ($P=2$, $D=256$, $H=4$). Note that most of our experiments were performed without residual connections in Eqs. \ref{eq:res} and \ref{eq:resff}. As described later in \ref{sec:numlayer}, adding residual connections further improved performance.

We used 1024 internal units in a position-wise feed-forward layer ($d_\mathrm{ff}=1024)$.
We used the Adam optimizer with the learning rate scheduler described in \cite{Vaswani2017}. The number of warm-up steps used in the learning rate scheduler was 25,000. The batch size was 64. The number of training epochs was 100.
After 100 epochs, we used an averaged model obtained by averaging the model parameters of the last ten epochs.
As with the BLSTM-EEND system, we applied 11-frame median filtering.

For domain adaptation, the averaged model was retrained using the CALLHOME adaptation set.
We used the Adam optimizer with a learning rate of $10^{-5}$ and ran 100 epochs.
After 100 epochs, we used an averaged model obtained
by averaging the model parameters of the last ten epochs.

\subsection{Performance metric}

We evaluated the systems with the diarization error rate (DER) \cite{NISTRT09}. Note that the DERs reported in many prior studies did not include misses or false alarm errors due to their using oracle speech/non-speech labels. Overlapping speech segments had also been excluded from the evaluation. For our DER computation, we evaluated all of the errors, including overlapping speech segments, because the proposed method includes both the speech activity detection and overlapping speech detection functionality. As is done typically, we used a collar tolerance of 250 ms at the start and end of each segment.

\section{Results}
\label{sec:results}

\begin{table}[t]
\caption{DERs (\%) on various test sets. For EEND systems, the CALLHOME (CH) results were obtained with domain adaptation.
}
\label{tab:overlap}
\centering
\begin{tabular}{l|ccc|cc} \hline
 & \multicolumn{3}{c|}{Simulated} & \multicolumn{2}{c}{Real} \\
 & $\beta=2$ & $\beta=3$ & $\beta=5$ & CH & CSJ \\ \hline 
Clustering-based & & & & &\\
\: i-vector & 33.74 & 30.93 & 25.96 & 12.10 & 27.99 \\
\: x-vector &    28.77 & 24.46 & 19.78 & 11.53 & 22.96 \\ \hline
BLSTM-EEND & & & & & \\
\: trained with SimBeta2 & 12.28 & 14.36 & 19.69 & 26.03 & 39.33 \\
\: trained with Real & 36.23 & 37.78 & 40.34 & 23.07 & 25.37 \\ \hline
SA-EEND & & & & &\\
\: trained with SimBeta2  & 7.91 & 8.51 & 9.51 & 13.66 & 22.31 \\
\: trained with Real  & 32.72 & 33.84 & 36.78 & {\bf 10.76} & \bf 20.50 \\
\: trained with SimLarge & {\bf 6.81} &  6.60 & 6.40 & 14.03 & 21.84 \\
\: trained with Comb  & 6.92 & {\bf 6.54} & \bf{6.38} & 11.99 & 22.26 \\
\hline
\end{tabular}
\end{table}

\begin{table}[t]
\caption{DERs (\%) on the CALLHOME with and without domain adaptation.
}
\label{tab:adapt}
\centering
\begin{tabular}{l|cc} \hline
 & w/o adaptation & with adaptatation \\ \hline
x-vector clustering & 11.53 & N/A \\ \hline
BLSTM-EEND \\
\: trained with SimBeta2 & 43.84 & 26.03 \\
\: trained with Real & 31.01 & 23.07 \\ \hline
SA-EEND \\
\: trained with SimBeta2 & 17.42 & 13.66 \\
\: trained with SimLarge & 16.31 & 14.03 \\
\: trained with Real & 12.66 & \bf{10.76} \\
\: trained with Comb & 14.50 & 11.99 \\ \hline
\end{tabular}
\end{table}

\begin{table}[t]
\caption{Detailed DERs (\%) evaluated on the CALLHOME. DER is composed of Misses (MI), False alarms (FA), and Confusion errors (CF). The SAD errors are composed of Misses (MI) and False alarms (FA) errors.}
\label{tab:detail}
\centering
\begin{tabular}{l|c|ccc|cc} \hline
    & & \multicolumn{3}{c|}{DER breakdown} & \multicolumn{2}{c}{SAD errors} \\
Method & DER & MI & FA & CF & MI & FA \\ \hline
i-vector & 12.10 & 7.74 & 0.54 & 3.82 & 1.4 & 0.5 \\
x-vector &    11.53 & 7.74 & 0.54 & 3.25 & 1.4 & 0.5 \\ \hline
\begin{tabular}[b]{@{}c@{}}SA-EEND\\\quad no-adapt\end{tabular} &    12.66 & 7.42 & 3.93 &  1.31 & 3.3 & 0.6 \\
\quad adapted  & {\bf 10.76} & 6.68 & 2.40 & 1.68 & 2.3 & 0.5 \\ \hline
\end{tabular}
\end{table}

\begin{figure*}[t]
\begin{center}
    \includegraphics[width=\linewidth]{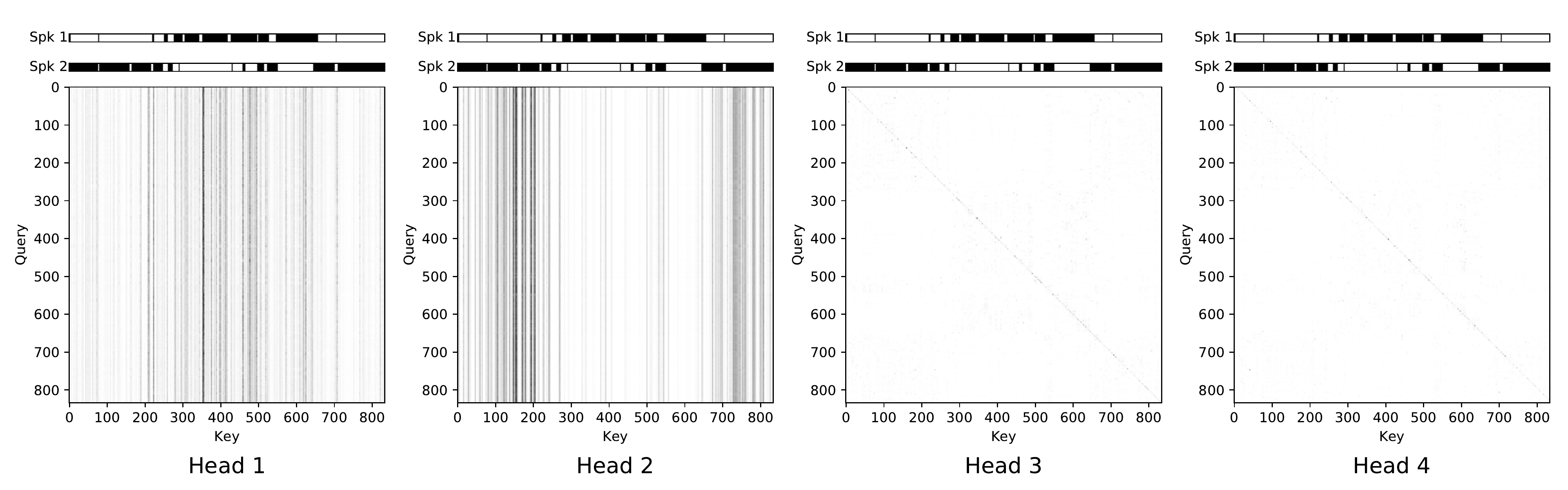}
\end{center}
\caption{Attention weight matrices at the second encoder block. The input was the CALLHOME test set (recording id: iagk). The model was trained with the real training set followed by domain adaptation. The top two rows show the reference speech activity of two speakers.}
\label{fig:visualize}
\end{figure*}


\subsection{Evaluation on simulated mixtures}

DERs on various test sets are shown in Table \ref{tab:overlap}.
The clustering-based systems performed poorly on heavily overlapping simulated mixtures.
This result is within our expectations because the clustering-based systems did not consider speaker overlaps; there were more misses when the overlap ratio was high.

The BLSTM-EEND system trained with the simulated training set (SimBeta2) showed a significant DER reduction compared with the clustering-based systems on the simulated mixtures.
Among the differing overlap ratios, it performed the best on the highest overlap ratio condition ($\beta=2$).
The BLSTM-EEND system worked well on the overlapping condition matched that of the training data.

The SA-EEND system trained with the simulated training set had significantly fewer DERs compared with the BLSTM-EEND system on every test set.
As well as the BLSTM-EEND system, it showed the best performance on the highest overlap ratio condition ($\beta=2$).
However, the DER degradation under fewer overlapping conditions was smaller than that of the BLSTM-EEND system, which indicated that the self-attention blocks improved robustness to variable overlapping conditions.

Training the SA-EEND model with various overlap ratio conditions (SimLarge) showed an improvement over the single overlap ratio condition (SimBeta2) on every test set. It was revealed that overfitting to a specific overlap ratio could be mitigated by this multi-condition training.

\subsection{Evaluation on real test sets}

In contrast to the excellent performance on the simulated mixtures, the BLSTM-EEND system had inferior DERs to those of the clustering-based systems evaluated on the real test sets.
Although the BLSTM-EEND system showed performance improvements when the training data were switched from simulated to real data, its DERs were still higher than those of the clustering-based systems. 

The SA-EEND system trained with the simulated training set (SimBeta2) showed remarkable improvements on the real test sets of CALLHOME and CSJ, which indicates the strong generalization capability of the self-attention blocks.
For the CSJ, even without domain adaptation, the SA-EEND system performed better than the x-vector clustering-based method.
Training the SA-EEND model with various overlap ratio conditions (SimLarge) yielded excellent generalizations to real test sets.

The SA-EEND system trained with the real training set (Real) performed better than SimLarge on the real test sets. However, it had poor DERs on the simulated test sets. We believed that the result was due to the small number of mixtures and low overlap ratio of the real training set. 
Finally, the SA-EEND system trained with the combined dataset (Comb) showed an excellent generalization capability,
which was obtained by feeding it various overlap ratio conditions.

\subsection{Effect of domain adaptation}

The EEND models trained with simulated training set were overfitted to the specific overlap ratio of the training set.
We expected that the overfitting would be mitigated by using domain adaptation.
DERs on the CALLHOME with and without domain adaptation are shown in Table \ref{tab:adapt}.
As expected, the domain adaptation significantly reduced the DER; our system thus achieved even better results than those of the x-vector-based system.

A detailed DER comparison on the CALLHOME test set is shown in Table \ref{tab:detail}. The clustering-based systems had few SAD errors thanks to the robust SAD model trained with various noise-augmented data. However, there were numerous misses and confusion errors due to its lack of handling speaker overlaps. Compared with clustering-based systems, the proposed method produced significantly fewer confusion and miss errors.
The domain adaptation reduced all error types except confusion errors.

\subsection{Visualization of self-attention}
\label{sec:vis}
To analyze the behavior of the self-attention mechanism in our diarization system, Fig. \ref{fig:visualize} visualizes the attention weight matrix at the second encoder block, corresponding to $\mathbf{\hat{A}}^{(p=2)}_h$ in Eq. \ref{eq:softmax}.
Here, head 1 and head 2 have vertical lines at different positions.
The vertical lines correspond to each speaker's activity.
The attention weight matrix with these vertical lines transformed the input features into the weighted mean of the same speaker frames. These heads actually captured the global speaker characteristics by computing the similarity between distant frames.
Interestingly, heads 3 and 4 look like diagonal matrices, which result in local linear transforms. These heads are considered to act as speech/non-speech detectors.
We conclude that the multi-head self-attention mechanism captures global speaker characteristics in addition to local speech activity dynamics, which leads to a reduction in DER.

\pdffigure[width=\linewidth]{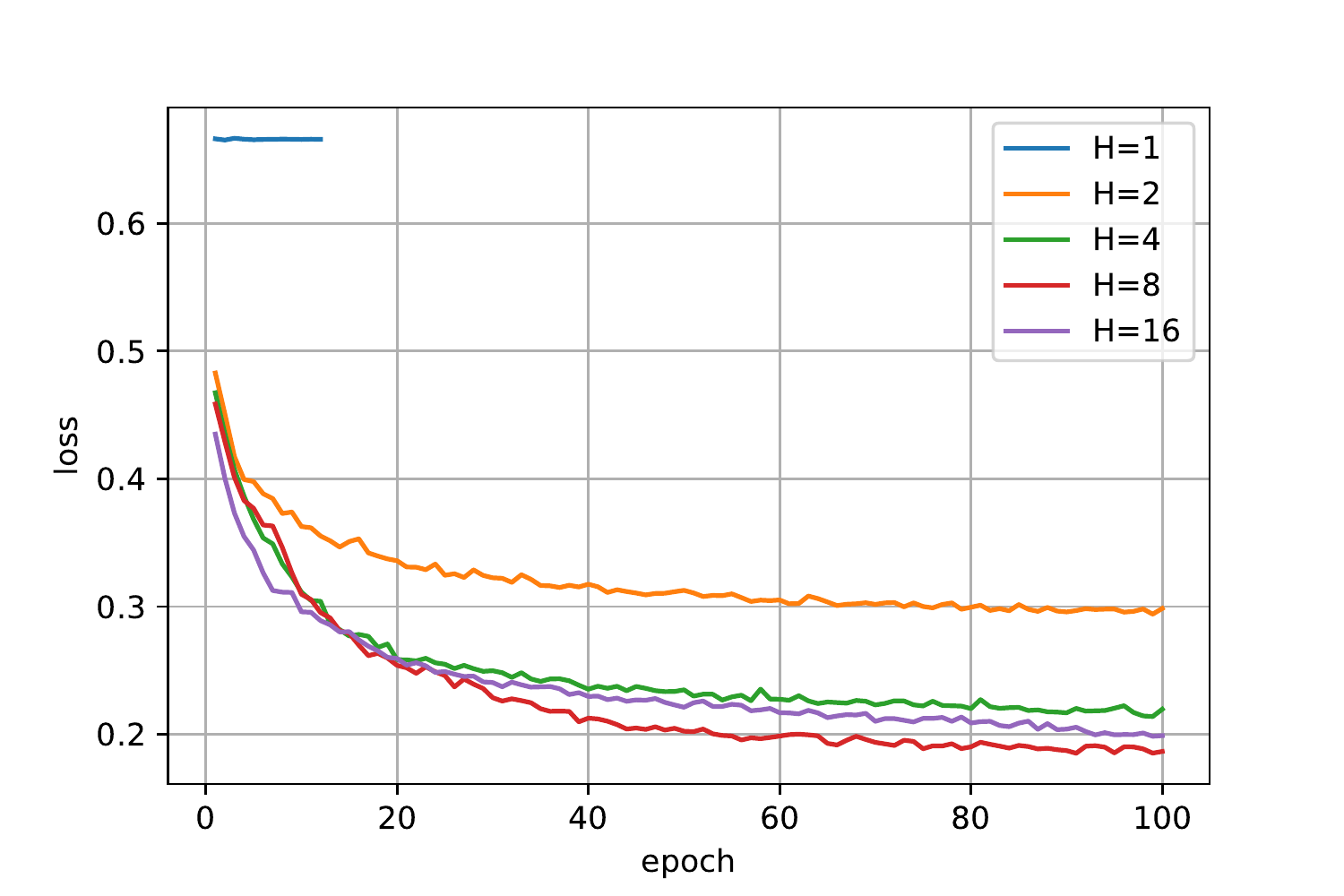}{Loss curves on simulated validation set ($\beta=2$) for different numbers of heads. These models were trained with SimBeta2.}

\subsection{Effect of varying number of heads in self-attention blocks}
\begin{table}[t]
\caption{DERs (\%) with different number of heads. The models are trained with SimBeta2.}
\label{tab:nh}
\centering
\begin{tabular}{r|rrr|rr} \hline
Num. & \multicolumn{3}{c|}{Simulated} & \multicolumn{2}{c}{Real} \\
heads & $\beta=2$ & $\beta=3$ & $\beta=5$ & CH & CSJ  \\ \hline
2 & 12.60 & 13.42 & 16.12 & 16.49 & 26.05  \\
4 & 7.91 & 8.51 & 9.51 & 13.66 & {\bf22.31} \\
8 & \bf6.84 & \bf7.06 & \bf7.85 & 13.44 & 23.58 \\
16 & 7.19 & 7.52 & 7.88 & {\bf13.28} & 24.35 \\ \hline
\end{tabular}
\end{table}
The analysis in Sec. \ref{sec:vis} indicated that the different heads represented different speakers. To verify the importance of multiple heads, we trained models with different numbers of heads. The loss curves with for those models are shown in Fig. \ref{fig:numheads_valid_loss}. The loss decreased as the number of heads increased and this trend continued for a large number of epochs. Note that for the single-head ($H=1$) experiment, we interrupted the training because the losses were consistent, around 0.67 during the first 12 epochs.

The DERs for different numbers of heads are shown in Table \ref{tab:nh}. Here, performance improved as a result of increasing the number of heads. These results suggest that the SA-EEND models were trained to separate speakers via the global speaker characteristics represented by different heads, the required number of heads was at least the number of speakers, and more heads boosted performance.

\subsection{Effect of varying number of encoder blocks and warm-up steps}
\label{sec:numlayer}
\pdffigure[width=\linewidth]{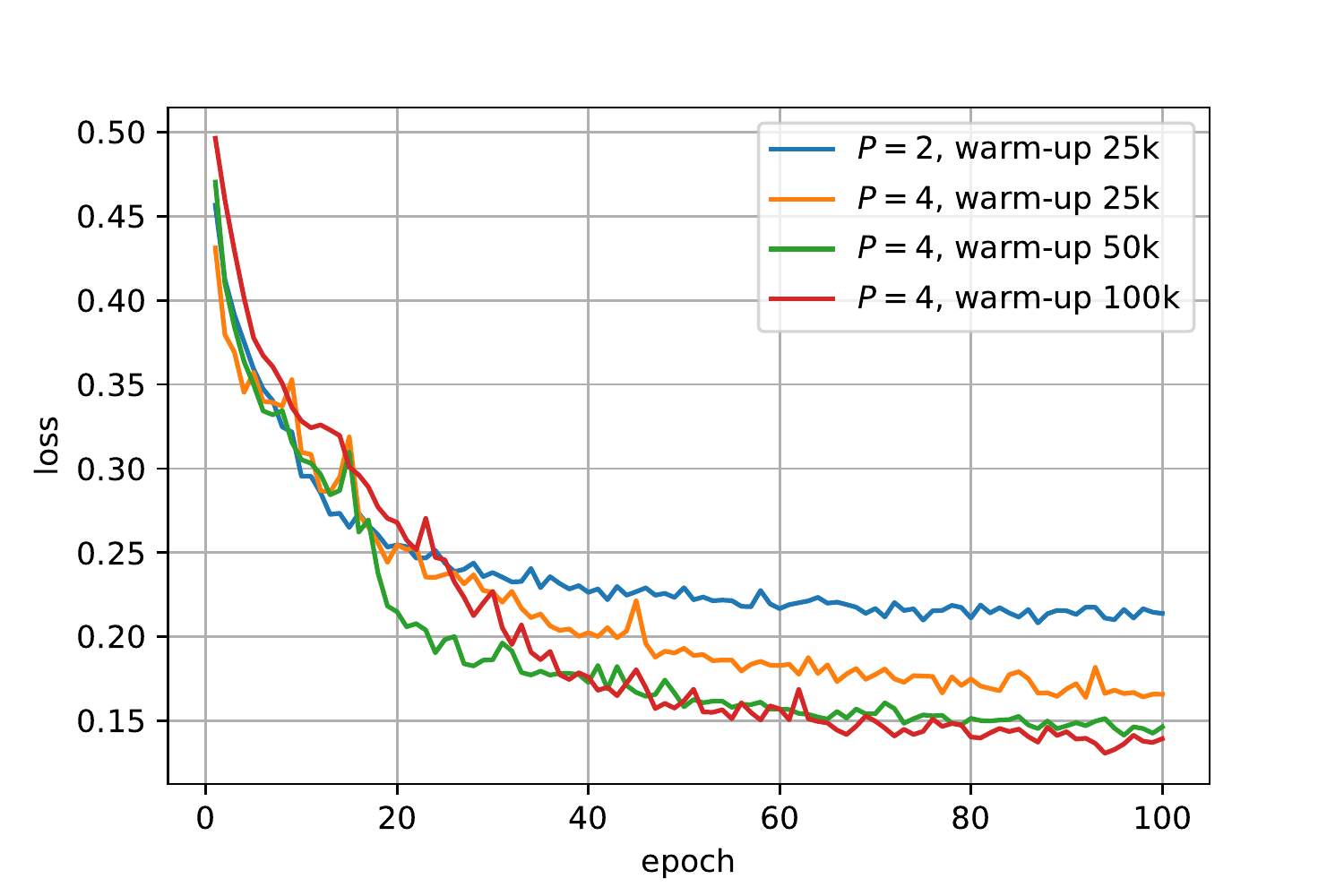}{Loss curves on simulated validation set ($\beta=2$) for different numbers of layers and warm-up steps. These models were trained with SimBeta2.}

\begin{table}[t]
\caption{DERs (\%) for different numbers of encoder blocks and warm-up steps with/without residual connections. The models were trained with SimBeta2}
\label{tab:nl}
\centering
\begin{tabular}{rrr|rrr|rr} \hline
Enc. & Warm. & Res. & \multicolumn{3}{c|}{Simulated} & \multicolumn{2}{c}{Real} \\
blocks & steps & con. & $\beta=2$ & $\beta=3$ & $\beta=5$ & CH & CSJ  \\ \hline
2 & 25k & N & 7.91 & 8.51 & 9.51 & 13.66 & 22.31 \\ 
2 & 25k & Y & 7.36 & 7.59 & 7.78 & 12.50 & 23.38  \\
4 & 25k & Y & 5.66 & 5.39 & 5.01 & 10.16 & {\bf 20.39}\\
4 & 50k & Y & 5.01 & 4.64 & 4.10 & 10.25 & 21.50\\
4 & 100k & Y & {\bf 4.56} & {\bf 4.50} & {\bf 3.85} & {\bf 9.54} & 20.48\\
\hline
\multicolumn{3}{c|}{x-vector clustering} & 28.77 & 24.46 & 19.78 & 11.53 & 22.96 \\ \hline

\end{tabular}
\end{table}

As noted in Sec. \ref{sec:eendsetup}, most of our experiments were performed without residual connections in Eqs. \ref{eq:res} and \ref{eq:resff}. In this section, we examined deeper model configurations using more encoder blocks with residual connections.
The loss curves for different numbers of encoder blocks and warm-up steps are shown in Fig. \ref{fig:numlayers_valid_loss}. The models with four encoder blocks reduced the validation loss compared with the one with two encoder blocks. Moreover, the validation loss was reduced by increasing the number of warm-up steps from 25,000 to 100,000.
DERs for different numbers of encoder blocks are shown in Table \ref{tab:nl}.
The results show that increasing the number of encoder blocks significantly improved performance.

The EEND system achieved a DER of 9.54\%, whereas the x-vector clustering-based system had a DER of 11.53\% on the CALLHOME dataset. Moreover, EEND had a DER of 20.39\% on the CSJ dataset, while the x-vector clustering-based system had 22.96\%. EEND had DERs from 4.56\% to 3.85\% on the simulated test set, while the x-vector clustering-based system had 19.78\% to 28.77\%.

\section{Conclusion}
\label{sec:print}

We proposed End-to-End Neural Diarization (EEND), in which a neural network directly outputs speaker diarization results given a multi-speaker recording.
We formulated the speaker diarization problem as a multi-label classification problem and introduced a permutation-free objective function to minimize diarization errors directly.
We evaluated our method on simulated speech mixtures and real conversation datasets.
The results showed that EEND method outperformed that of the state-of-the-art x-vector clustering-based method, and it correctly handled speaker overlaps.
We explored the neural network architecture for the EEND method, and found that the self-attention-based neural network was the key to achieving excellent performance.
By visualizing the attention weights, we showed that self-attention captured the global speaker characteristics in addition to local speech activity dynamics, making it especially suitable for dealing with the speaker diarization problem.
Experiments with different numbers of heads showed that the excellent performance could be obtained by making the number of heads sufficiently larger than the number of speakers.
Finally, experiments with different numbers of encoder blocks revealed that the EEND model performed better when it had more encoder blocks.


\ifCLASSOPTIONcaptionsoff
  \newpage
\fi



\bibliographystyle{IEEEtran}
\bibliography{refs}
%

%

\begin{IEEEbiography}{Yusuke Fujita}
received the B.S. and M.S. degree in computer science from Waseda University, Tokyo, Japan, in 2003 and 2005, respectively.
Currently, he is a Senior Researcher of Media Intelligent Systems Research Department, Hitachi, Ltd., Tokyo, Japan.
He is also a Visiting Scholar at Johns Hopkins University, MD, USA.
His research interests include speech recognition, speech separation and speaker diarization.
\end{IEEEbiography}

\begin{IEEEbiography}{Shinji Watanabe} is an Associate Research Professor at Johns Hopkins University, Baltimore, MD, USA. 
He received his B.S., M.S., PhD (Dr. Eng.) Degrees in 1999, 2001, and 2006, from Waseda University, Tokyo, Japan. 
He was a research scientist at NTT Communication Science Laboratories, Kyoto, Japan, from 2001 to 2011, a visiting scholar in Georgia institute of technology, Atlanta, GA in 2009, and a Senior Principal Research Scientist at Mitsubishi Electric Research Laboratories (MERL), Cambridge, MA from 2012 to 2017.
His research interests include automatic speech recognition, speech enhancement, spoken language understanding, and machine learning for speech and language processing. 
He has been published more than 200 papers in top journals and conferences, and received several awards including the best paper award from the IEICE in 2003. 
He served an Associate Editor of the IEEE Transactions on Audio Speech and Language Processing, and is a member of several technical committees including the IEEE Signal Processing Society Speech and Language Technical Committee (SLTC) and Machine Learning for Signal Processing Technical Committee (MLSP).
\end{IEEEbiography}


\begin{IEEEbiography}{Shota Horiguchi}
received the B.S. degree in information and communication engineering and the M.S. degree in information science and technology from the University of Tokyo, Tokyo, Japan, in 2015 and 2017, respectively.
He is currently a Researcher with Hitachi, Ltd, Tokyo, Japan.
\end{IEEEbiography}

\begin{IEEEbiography}{Yawen Xue}
Biography text here.
\end{IEEEbiography}

\begin{IEEEbiography}{Nagamatsu Kenji}
Biography text here.
\end{IEEEbiography}




\end{document}